\documentclass{article}
\usepackage[utf8]{inputenc}
\usepackage{amsmath}
\usepackage{amssymb}
\usepackage{bbm}
\usepackage{todonotes}
\usepackage{amsthm}
\usepackage{hyperref}
\newcommand{\email}[1]{\href{mailto:#1}{#1}}

\newcommand{\Z}{\mathbb Z}
\newcommand{\R}{\mathbb R}

\newcommand{\Norm}[1]{\left\| #1 \right\|}

\newcommand*{\tran}{^{\mkern-1.5mu\mathsf{T}}}

\newtheorem{theorem}[]{Theorem}
\newtheorem{corollary}[theorem]{Corollary}

\title{Optimizing Low Dimensional Functions \\ over the Integers\thanks{The fourth author was partially funded by Fondecyt grant Nr. 1221460 and Centro de Modelamiento Matemático (CMM), FB210005, BASAL funds, ANID-Chile.}}

\author{Daniel Dadush\footnote{CWI, Amsterdam, Netherlands, \email{dadush@cwi.nl}} \and Arthur L\'eonard\footnote{ENS, Paris, France, \email{arthur.leonard@ens.psl.eu}} \and Lars Rohwedder\footnote{Maastricht University, Maastricht, Netherlands, \email{l.rohwedder@maastrichtuniversity.nl}} \and Jos\'e Verschae\footnote{Pontificia Universidad Cat\'olica de Chile, Santiago, Chile, 
\email{jverschae@uc.cl}}}

\date{}

\begin{document}

\maketitle

\begin{abstract}
We consider box-constrained integer programs with objective $g(Wx) + c\tran x$, where $g$ is a ``complicated'' function with an $m$ dimensional domain. Here we assume we have $n \gg m$ variables and that $W \in \Z^{m \times n}$ is an integer matrix with coefficients of absolute value at most $\Delta$. We design an algorithm for this problem using only the mild assumption that the objective can be optimized efficiently when all but $m$ variables are fixed, yielding a running time of $n^m(m \Delta)^{O(m^2)}$. Moreover, we can avoid the term $n^m$ in several special cases, in particular when $c = 0$. 

Our approach can be applied in a variety of settings, generalizing several recent results. An important application are convex objectives of low domain dimension, where we imply a recent result by Hunkenschr\"oder et al.~[SIOPT'22] for the 0-1-hypercube and sharp or separable convex~$g$, assuming $W$ is given explicitly. By avoiding the direct use of proximity results, which only holds when $g$ is separable or sharp, we match their running time and generalize it for arbitrary convex functions. In the case where the objective is only accessible by an oracle and $W$ is unknown, we further show that their proximity framework can be implemented in $n (m \Delta)^{O(m^2)}$-time instead of $n (m \Delta)^{O(m^3)}$. Lastly, we extend the result by Eisenbrand and Weismantel [SODA'17, TALG'20] for integer programs with few constraints to a mixed-integer linear program setting where integer variables appear in only a small number of different constraints.
\end{abstract}

\section{Introduction}
Integer programming has played a crucial role in many areas of computer science, operations research, and more recently, data science. Its modelling power allows to capture a large diversity of settings. However, its general intractability makes it challenging to derive a general algorithmic theory, and hence the focus has been to consider meaningful special cases. The main theoretical result in this area has been the algorithm by Lenstra~\cite{lenstra1983integer}, and the improvement by Kannan~\cite{kannan1983improved}, which show that integer programs are tractable as long as the dimension is constant. In recent years, a surge of interest appeared regarding efficient algorithms for integer programs under other assumptions. More recently, the seminal work by Eisenbrand and Weismantel~\cite{EisenbrandW2019} for integer programs with a constant number of constraints and bounded matrix coefficients sparked a new trend of improved algorithms and lower bounds; see, e.g.,~\cite{intprogconv,KnopPW20,BrandKO21}. 

In this paper, we study a new general framework that encompasses and further extends many of the settings found in the literature. Consider the problem of optimizing a low dimensional objective function over a high dimensional space~$\Z^n$. Formally, the problem is defined as
\begin{align}
    \min\ &c\tran x + g(W x) \notag \\
    & \ell_i \le x_i \le u_i \label{eq:problem} & \text{for all } i\in\{1,2,\dots,n\}, \\
    &x \in \Z^n. \notag
\end{align}
We assume that $W \in \Z^{m \times n}$ has entries of absolute value at most $\Delta$.
Here, $W$ can be interpreted as a projection matrix to a space of low dimension $m\ll n$, where then the function $g: \R^m \rightarrow \R \cup \{\infty\}$ is applied to the projection. We can think of $W$ as extracting a relatively
small set of features from $x$. The vectors $\ell \in (\Z\cup \{-\infty\})^n$ and $u\in (\Z \cup \{\infty\})^n$ are arbitrary variable bounds
and $c$ represents a linear cost function. 

Crucially, we only make a very mild assumption
on $g$, namely that we can solve~\eqref{eq:problem} when all but $m$ of the variables are fixed:
given any $I \dot\cup J = [n]$ with $|I| = m$ and any fixing $z \in \Z^J$ of the $J$-variables, we require that 
\begin{align}
    \min \quad & c_I x + g(W_I x +W_J z) \notag \\
    \text{s.t. }\quad &\ell_i \le x_i \le u_i &\text{for all } i\in I, \label{eq:query} \\
    & x \in \Z^I, \notag
\end{align}
can be solved efficiently. Here $c_I=(c_i)_{i\in I}
$ is the vector $c$ restricted to indices in $I$
and similarly $W_I$ (resp. $W_J$) is the matrix $W$ restricted to
columns indexed by $I$ (resp. $J$).
The requirement is intuitively necessary, because the only plausible
approach to efficiently solve the very general setting~(\ref{eq:problem})
is to exploit that the function $g$ is low dimensional.
If we cannot even optimize over it in a low dimensional coordinate subspace of $\Z^n$,
then there is no hope to optimize over it on all of $\Z^n$.
Perhaps the most natural such setting is
when $g$ is convex and can be accessed through gradient and function evaluation queries. Then~(\ref{eq:query}) can be solved in time that is exponential only
in $m$, but polynomial in the other input parameters, by the Lenstra-Kannan algorithm~\cite{lenstra1983integer,kannan1983improved}.

If $g$ is indeed convex, the Lenstra-Kannan algorithm can also be used to directly solve~\eqref{eq:problem} in time
$\Delta^{\Delta^{O(m)}} \cdot \langle\mathrm{input}\rangle^{O(1)}$, where $\langle\mathrm{input}\rangle$ denotes the encoding size of the input. Indeed, we can merge variables with the same columns in $W$, which reduces the dimension of the problem
to $n' = (2\Delta + 1)^m$. Notice that the linear part of the objective may not remain
linear, but it does remain convex. Thus, we can apply Lenstra-Kannan to solve the problem in the claimed running time.

This indicates that the problem is tractable for small values of $\Delta$ and $m$.
Our main result is an algorithm that avoids the double exponential running~time.

\begin{theorem}\label{thm:main}
For any function $g$, problem~(\ref{eq:problem}) can be solved in time
\begin{equation*}
    n^m \cdot (m \Delta)^{O(m^2)} \cdot Q \ ,
\end{equation*}
where $Q$ is the query time of the oracle for~(\ref{eq:query}). In particular, for a convex
function $g$, the term $Q$ can be replaced by $\langle\mathrm{input}\rangle^{O(1)}$.
\end{theorem}
We notice that in this theorem the bound of $Q$ for the convex case follows by using the Lenstra-Kannan algorithm to solve the small dimensional subproblem~\eqref{eq:query}. In this case, the $m^{O(m)}$ factor in the running time of the Lenstra-Kannan algorithm can be omitted as it is upper bounded by $(m \Delta)^{O(m^2)}$. Regarding the $n^m$ term, as we explain below, it can be made lower order in interesting concrete settings. We also remark that a term of the form $\Delta^m$ cannot be avoided due to reductions from integer linear programming (see Section~\ref{sec:applications}) and lower bounds for that problem~\cite{intprogconv}.

\subsection{Applications}\label{sec:applications}
\paragraph*{Low dimensional convex functions.}
The main inspiration for this work is a recent study by Hunkenschr\"oder, Pokutta, and Weismantel~\cite{hunkenschroder2022optimizing}, who consider the problem
\begin{equation}\label{eq:problem-hunk}
    \min_{x\in \{0,1\}^n} g(W x) \ ,
\end{equation}
where $W\in \Z^{m \times n}$ with entries of absolute value at most $\Delta$, $g: \R^m \rightarrow \R$ is a ``nice'' \emph{sharp} or separable convex function, and the algorithm can make function and gradient evaluations to the objective $g(Wx)$. They further distinguish between the case where $W$ is given explicitly and where $W$ is unknown to the algorithm. Assuming $g$ is separable, they provide an $n (m \Delta)^{O(m^2)}$-time algorithm when $W$ is known, and an $n (m L \Delta)^{O(m^3)}$-time algorithm when $W$ is unknown and $g$ is assumed to have $L$-Lipschitz gradients\footnote{They further require $g$ to have an integer valued gradient on integer inputs.}. They show similar results when $g$ is suitably ``sharp'', though we omit the statements for concision.

As a direct application of Theorem~\ref{thm:main}, we extend the result of~\cite{hunkenschroder2022optimizing} to arbitrary convex functions when $W$ is known. 
\begin{corollary}\label{cor:hunk}
When $W$ is known and $g$ is an arbitrary convex function, problem~(\ref{eq:problem-hunk}) can be solved in time
  \begin{equation*}
      O(nm) + (m \Delta)^{O(m^2)} \ .
  \end{equation*}
\end{corollary}
The reduction to Theorem~\ref{thm:main} is as follows: we first abandon the restriction of $x\in\{0,1\}^n$
in favor of the general bounded integer variables. Then, since any two variables with the same column in the projection matrix $W$ can be merged to one (by adapting the box-constraints), we may assume without loss of generality that $n \le (2\Delta + 1)^m$.

One of the main motivations in the work by Hunkenschröder et al.~\cite{hunkenschroder2022optimizing} is to solve certain types of regression problems. For example, they examine an integer \emph{compressed sensing} problem, where one receives a small number $m$ of linear measurements of a high dimensional integral signal $x^*\in\{0,1\}^n$ which one would like to (approximately) reconstruct. The received measurements are of the form $b = W x^*$, where $W \in \Z^{m \times n}$ is an unknown matrix with coefficients of size at most $\Delta$. As an approximation to $x^*$, they compute the minimizer of $\min\{\lVert b-Wx\lVert^2:x\in \{0,1\}^n\}$, under the assumption that one can only access~$W$ indirectly via gradient and function evaluation queries to $f(x) = \lVert b-Wx\lVert^2$.

As we will explain later, in the compressed sensing and related settings, one can essentially avoid any overhead from not knowing $W$. While we focus above on the case where $W$ is known, using orthogonal techniques, we can also improve the running times in the unknown $W$ setting by modifying the Hunkenschröder et al.\ framework. We defer further discussion of their framework and our related improvements to Section~\ref{sec:unknownW}.

\paragraph*{Mixed-integer linear programming.}
Eisenbrand and Weismantel~\cite{EisenbrandW2019}
studied the complexity of integer programs of the form
\begin{align}
    \min\ &c\tran x \notag\\
     \text{s.t. } &Ax = b,\label{eq:EW} \\
    & \ell_i \le x_i \le u_i &\text{for all }i\in\{1,\ldots,n\},\notag\\
     &x \in \Z^n.\notag
\end{align}
Specifically, they considered the setting where $A$ has few rows and then used the Steinitz
Lemma to obtain an algorithm with running time
$(m \Delta)^{O(m^2)} \cdot n$, where $\Delta$ is the maximum absolute value in $A$.
This has inspired a line of work for similar
settings, see for example~\cite{intprogconv,KnopPW20,CslovjecsekEHRW21,Klein22}.
Our setting is
a generalization of theirs: take $A = W$ and let
\begin{equation*}
    g(Ax) = \begin{cases}
      0 &\text{ if } Ax = b , \\
      \infty &\text{ otherwise.}
    \end{cases}
\end{equation*}
Here subproblem~\eqref{eq:query} corresponds to solving integer programming in $m$ dimensions,
which can be done using Lenstra-Kannan.
Alternatively, one could model the problem as
minimizing the convex function $g(Ax) = \Norm{Ax - b}$ for some suitable norm.
Moreover, our model generalizes beyond the scope of Eisenbrand and Weismantel's work
to mixed-integer linear programming. Consider
the problem
\begin{align}
    \min\ &c\tran x + d\tran y \notag \\
    \text{s.t. } &A x + B y = b, \notag \\
    &\ell_i \le \ x_i \le u_i, \label{eq:MILP}&\text{for all }i\in\{1,\ldots,n\} \\
    &x \in \Z^n, \notag \\
    &y \in P \subseteq \R^h.\notag
\end{align}
Here $P$ is some polytope that can impose additional constraints on the continuous variables.
We can encode this problem in~(\ref{eq:problem}) by setting $W = A$ and
\begin{equation*}
    g(Ax) = \begin{cases}
      \min\{d\tran y : By = b - Ax, y\in P\} &\text{if this minimum exists}, \\
      \infty &\text{otherwise.}
    \end{cases}
\end{equation*}
Notice that the oracle problem~(\ref{eq:query}) in this case forms a
mixed-integer linear program itself, but with only $m$ many integer variables;
hence it can be solved efficiently with the algorithm by Lenstra-Kannan.
\begin{corollary}
  Assuming $P$ can be efficiently separated over, problem~(\ref{eq:MILP}) can be solved in time
  \begin{equation*}
      n^m \cdot (m \Delta)^{O(m^2)} \cdot \langle\mathrm{input}\rangle^{O(1)} \ .
  \end{equation*}
\end{corollary}
We emphasize here that $\Delta$ is only a bound on the entries of $A$, but not necessarily on those of $B$.
Compared to the algorithm for the pure integer setting in~\cite{EisenbrandW2019}, our running time has an extra
factor of $n^m$, which however vanishes in some settings:
for example, when $c = 0$ or $u_i = \infty$ for all $i$.
In those cases we can again merge variables that share the same column in $W$.
The only other example we are aware of that extends Eisenbrand and Weismantel's setting to mixed-integer
linear programming is the work by Brand, Kouteck\'y, and Ordyniak~\cite{BrandKO21}.
Their setting can be considered orthogonal to ours. On the one hand, they study a much more general
structure of bounded treedepth programs, of which integer programs with a bounded number of constraints
are the simplest special case. On the other hand, they impose these structural restrictions also on the continuous variables (and additionally bounds on their coefficients),
whereas we impose essentially no restrictions on the structure of continuous variables or their coefficients.

To appreciate this, let us remark a pleasing aspect of the (straight-forward) 
extension of Lenstra-Kannan to mixed-integer linear programs:
it combines the tractability of integer programs in fixed
dimension with the tractability of linear programs in any dimension,
achieving essentially a generalization of both.
Eisenbrand and Weismantel's algorithm, on the other hand, concerns the tractability of integer programs with
a fixed number of constraints (adding the necessary assumption
that $\Delta$ is bounded). In a similar spirit to the
aforementioned generalization, our algorithm combines this with
the tractability of (arbitrary) linear programs.

\paragraph*{Integer linear programming with few complex variables.}
Recall the integer programming setting~\eqref{eq:EW}
studied by Eisenbrand and Weismantel, for which they
gave an algorithm with running time $(m\Delta)^{O(m^2)} \cdot n$ (with $\Delta$ being the maximum absolute value in $A$).
The interesting parameter regime for this algorithm
is therefore when $m$ and $\Delta$ are very small.
Already for $m = 1$ this formulation easily captures the Knapsack problem, which is weakly NP-hard and therefore we cannot hope to reduce the dependency on $\Delta$ to, say, $\log(\Delta)$ while still maintaining polynomial dependency on $n$.
In Lenstra-Kannan, on the other hand, the dependency on the coefficients of the matrix is polynomial in the encoding size, i.e., in $\log(\Delta)$, but the dependency on $n$ is exponential.
These two rather orthogonal results can be combined using
Theorem~\ref{thm:main}.
\begin{corollary}
    Consider the integer programming problem in~\eqref{eq:EW}
    and partition the columns of $A$ into ``simple'' columns
    where the entries are bounded by $\Delta$ in absolute value and ``complex'' columns where they are arbitrary.
    Suppose that there are only $k$ many complex columns. Then we can solve~\eqref{eq:EW} in time
    \begin{equation*}
        n^m \cdot (m\Delta)^{m^2} \cdot k^{O(k)} \cdot \langle\mathrm{input}\rangle^{O(1)} \ .
    \end{equation*}
\end{corollary}
For this we proceed as follows.
Let $S$ and $C$ be the index sets of the simple and complex columns and accordingly let $A_S$ and $A_C$ be the matrix $A$ restricted to these column sets.
We define Problem~\eqref{eq:problem} only on $x_S$, the variables for the simple columns. Then let

\begin{equation*}
    g(A_S x_S) = \begin{cases}
        \min\{c_C\tran x_C : A_C x_C = b - A_S x_S\} &\text{ if this minimum exists,}\\
        \infty &\text{ otherwise.}
    \end{cases} 
\end{equation*}
The resulting subproblem~\eqref{eq:query} is then an integer program with $m + k$ variables that can be solved using Lenstra-Kannan. We note that one could even add to~\eqref{eq:EW} arbitrary additional constraints on the complex columns
and still solve the problem in the same way.
\paragraph*{Variable-sized Knapsack.} Antoniadis et al.~\cite{antoniadis2013pack} introduce a variant of the Knapsack problem
with a non-linear cost function associated with the
used capacity. They show that the case where this function is concave is polynomial time solvable and describe an FPTAS for the convex case.
Our result can be used to devise a pseudopolynomial time
algorithm for the convex case: the problem can be expressed as
\begin{equation}\label{eq:knapsack}
    \max\left\lbrace \sum_{i=1}^n p_i x_i - g\bigg(\sum_{i=1}^n w_i x_i\bigg):\, x_i \in \{0,1,\dotsc,u_i\}\text{ for all } i\right\rbrace \ .
\end{equation}
where $p_i$ is the profit of item $i$, $w_i$ the weight, and $u_i\in\Z_{\ge 0}\cup \{\infty\}$
is a bound on the number of items of this type. Straightforward generalizations to multi-dimensional knapsack follow in a similar way.
\begin{corollary}
  For a convex function $g$, problem~(\ref{eq:knapsack}) can be solved in time
  \begin{equation*}
      (n + w_{\max})^{O(1)} .
  \end{equation*}
\end{corollary}
Here, the oracle problem~(\ref{eq:query}) reduces to a simple binary search.
In general our result fits well to problems with a similar spirit, where
the constraints are not hard, but they induce some penalty.

\subsection{Overview of Techniques}
\label{sec:techniques}

The related results for more restrictive
cases in~\cite{EisenbrandW2019}
and~\cite{hunkenschroder2022optimizing}
are based on proximity:
the continuous relaxation of the problem, where
the integer requirement is omitted, is solved
and if one can show that the solution
for the relaxation and the actual solution differ
only slightly, then this can be 
exploited in reducing the search space.
The precise proximity theorem in~\cite{EisenbrandW2019}
is as follows.
\begin{theorem}[Eisenbrand and Weismantel~\cite{EisenbrandW2019}]
\label{prop:prox}
  Let $z$ be an optimal vertex solution to the linear program
  \begin{equation*}
      \max\left\lbrace c\tran x: Ax = b \text{\rm{} and } \ell_i\le x_i\le u_i \text{\rm{} for all }i\right\rbrace,
  \end{equation*}
  where $A\in\Z^{m\times n}$ has entries of size at most $\Delta$.
  If there exists an integer solution, then there is also an optimal integer solution $x^*$ with
  \begin{equation*}
      \Norm{x^* - z}_1 \le m (2m \Delta + 1)^m \ .
  \end{equation*}
\end{theorem}

Hunkenschr\"oder et al.~\cite{hunkenschroder2022optimizing} consider the optimal solution to the
continuous relaxation of~(\ref{eq:problem-hunk}).
In the special cases of separable convex and strict convex functions they show that a similar proximity holds, which is a crucial ingredient in their algorithm.

Already for general convex functions, however, the proximity bound can be very large, as shown in an example in~\cite{hunkenschroder2022optimizing}.
This forms a serious obstacle towards our main
result.
We manage to circumvent this and still rely on proximity by applying it in a different way.
Consider for sake of illustration that we were able to determine the value of $b^* = W x^*$,
where $x^*$ is the optimal solution of~(\ref{eq:problem}). Then it would be easy to recover $x^*$ (solving
our problem) by applying
the integer linear programming algorithm by Eisenbrand and Weismantel~\cite{EisenbrandW2019}. The algorithm works
by computing the continuous solution $z$
to $W x = b^*$ and then using that
$\Norm{z - x^*}_1$ is bounded by Theorem~\ref{prop:prox}. Indeed, this bound still holds in our case
when fixing $b^*$.
However, it is not clear how to compute or guess $b^*$, nor how to compute $z$ without knowing $b^*$.

Let us now consider the case that the domain of each variable is $\Z_{\ge 0}$,
which is slightly simpler than the bounded case.
Here we may assume that $z$ has only $m$ non-zero components, which we
can guess from $n^m$ candidates. We still do not know $b^*$ or $z$, but we trivially know
$z$ on the $n-m$ zero components. Intuitively, this is enough to
apply proximity to recover $x^*$ on the zero components of $z$.
Moreover, recovering $x^*$ on the non-zero components of $z$ is only an $m$-dimensional
problem, where we can apply the oracle problem~(\ref{eq:query}).

For our general result with arbitrarily bounded variable domains, there is another obstacle: if we try to generalize the previous line of arguments,
it is still true that there are only $m$ ``special'' variables in $z$, namely variables that are not tight on either of their bounds. For the remaining variables,
however, it is not immediately obvious whether they equal the lower bound or the
upper bound and if we do not know this, it is unclear how to determine
$x^*$ on these tight variables. We overcome this by 
guessing enough information about the dual so that we can use
complementary slackness to infer which bound that the tight variables attain.

\section{Non-Negative Variables}
For simplicity, in this section we first prove
our main result
for the variable domain $\Z_{\ge 0}$, that is,
$\ell_i = 0$ and $u_i = \infty$
for all $i$.  
Consider the optimal solution $x^*$ to~(\ref{eq:problem}) and define
$b^* = W x^*$. Furthermore, let $z$ be an optimal vertex solution to
$\min\{c\tran x: W x = b^*, x\in \R_{\ge 0}^n\}$. We emphasize that $z$ is not necessarily integral.
By Theorem~\ref{prop:prox}
there is an optimal integer
solution $x'$ to $\min\{ c\tran x: W x = b^*, x\in \Z_{\ge 0}^n\}$ with
$\Norm{x' - z}_1 \le O(m \Delta)^m$. We can assume without loss
of generality that $x' = x^*$.
Since $z$ is a vertex solution, it has at least $n - m$ zero components $T$.
It follows that
\begin{equation*}
    \Norm{x^*_T}_1 = \Norm{x^*_T - z_T}_1 \le \Norm{x^* - z}_1 \le O(m \Delta)^m \ .
\end{equation*}
Thus, 
\begin{equation*}
    \Norm{W_T x^*_T}_1 \le m \Delta \cdot \Norm{x^*_T}_1 \le O(m \Delta)^{m+1} \ .
\end{equation*}
We now guess the indices of variables in $T$ from the $n^m$ many candidates and we guess
the value of $b^{(T)} := W_T x^*_T$ from the $O(m \Delta)^{(m+1)m}$ many candidates. It is now easy to recover $x^*_T$ (or an equivalent solution)
by solving
\begin{equation*}
    \min\left\lbrace c_T\tran x_T: W_T x_T = b^{(T)} \text{ and } x_i \in \Z_{\ge 0} \text{ for all } i\in T\right\rbrace.
\end{equation*}
Here we use the algorithm by Eisenbrand and Weismantel~\cite{EisenbrandW2019} or
the improvement in~\cite{intprogconv}.
This requires time $(m \Delta)^{O(m)} \cdot n$, which is insignificant compared to the number of guesses above. The algorithm assumes a solution to the LP relaxation is given, 
which, however, only serves the purpose of having a vector close to the optimal solution (in $\ell_1$-norm). For this purpose we can also simply take $z_T$ (the zero vector). 
Let $L$ be the set of indices not in $T$. To recover $x^*_L$ we need to solve
\begin{equation*}
    \min\left\lbrace c_L\tran x^*_L + g(W_L x^*_L + W_T x^*_T): x^*_i \in \Z_{\ge 0} \text{ for all } i\in L \right\rbrace.
\end{equation*}
This corresponds to an oracle query of the form~(\ref{eq:query}).
For each guess of $T$ and~$b^{(T)}$ we compute a solution in
this way and return the best among them.
The running time, which is dominated by the number of guesses, is therefore
\begin{equation}
    n^m \cdot (m\Delta)^{O(m^2)} \cdot Q \ .
\end{equation}
In fact, the $n^m$ term here can be omitted, since one may assume without loss of generality that
no two columns of $W$ are equal and therefore $n \le (2\Delta + 1)^m$.
\section{Bounded Variables}
Let again $x^*$ denote an optimal solution to~(\ref{eq:problem}) and $b^* = W x^*$.
Let $z$ be an optimal solution to
\begin{equation*}
    \min\left\lbrace c\tran x: Wx = b^* \text{ and } \ell_i \le x_i \le u_i \text{ for all } i \right\rbrace.
\end{equation*}
We assume that $c$ is augmented
slightly by adding $\varepsilon^i$ to the $i$th component for all $i$ for some very small $\varepsilon$, which essentially implements a lexicographic tie-breaking rule between solutions.
Here $\varepsilon$ can be treated symbolically.
We note that the dual of this linear program is
\begin{equation*}
    \max\left\lbrace {b^*}\tran y + \ell\tran s^{\ell} - u\tran s^u : c - W\tran y = s^{\ell} - s^{u} \text{ and } s^{\ell}_i, s_i^{u} \in \mathbb R_{\ge 0}^n, y\in \mathbb R^m \right\rbrace.
\end{equation*}
Let $y, s^{\ell}, s^{u}$ be an optimal vertex solution to the dual. Then there are $m$ linearly independent rows $(W\tran)_i$
with $s^{\ell}_i = s^{u}_i = 0$ (otherwise it would not be a vertex solution). We guess these rows among the $n^m$ candidates, which fully determines $y$ and in particular $c - W\tran y$.
We may assume that $(c - W\tran y)_i \neq 0$
for the other $n-m$ rows, which follows from the perturbation with $\varepsilon$.
If $(c - W\tran y)_i > 0$ for some $i$ we know
that $s^{u}_i > 0$ and likewise if $(c - W\tran y)_i < 0$, then
$s^{\ell}_i > 0$. By complementary slackness we can determine for these
rows that $z_i = u_i$ (respectively, $z_i = \ell_i$).
It follows that for $n - m$ variables $T$ we now determined
its value in $z$. Let $L$ denote the $m$ other variables.
We now proceed similar to the previous section.
We again have that
\begin{equation*}
    \Norm{x^*_T - z_T}_1 \le O(m\Delta)^m .
\end{equation*}
This implies that
\begin{equation*}
    \Norm{W_T x^*_T - W_T z_T}_1 \le m \Delta \cdot \Norm{x^*_T} \le O(m\Delta)^{m+1} \ .
\end{equation*}
Since we know the value of $W_T z_T$, we can guess $b^{(T)} = W_T x^*_T$ among
the $O(m\Delta)^{(m+1)m}$
many candidates.
Then we recover $x^*_T$ using the algorithm
by Eisenbrand and Weismantel~\cite{EisenbrandW2019} (where we can use $z_T$ instead of
an LP solution)
and $x^*_L$ by applying~(\ref{eq:query}) to
\begin{equation*}
    \min\{ c_L\tran x^*_L + g(W_L x^*_L + W_T x^*_T):  x^*_i \in \{\ell_i, \ell_i+1,\dotsc,u_i\} \text{ for all } i\in L \}.
\end{equation*}
Finally, we return the best solution computed for any guess.

\section{Overview of Hunkenschr\"oder et al.~\cite{hunkenschroder2022optimizing} and Related Improvements}
\label{sec:unknownW}

We now explain the high-level algorithm of Hunkenschröder et al.~\cite{hunkenschroder2022optimizing} in more detail, as well as some improvements to their framework in the unknown $W$ case.

Their algorithm starts with an optimal solution $z$ to the continuous relaxation $\min\{ g(Wx): x \in [0,1]^n\}$ having at most $m$ fractional components (which is easy to show to always exists). Here, $z$ is assumed to be given by an oracle. For the cases they consider, e.g., the separable case, they prove that there is a ``nearby'' optimal integral solution $x^*$ satisfying $\|x^*-z\|_1 \leq (m \Delta)^{O(m)}$.
Function $g$ being separable means that it can be decomposed into a sum of functions each depending only on a single dimension, that is, $g(Wx) = g_1((W x)_1) + \dotsc + g_m((W x)_m)$.
Using the proximity result, they guess $b^* = Wx^* \in \Z^m$, where the number of guesses is bounded by $(m \Delta)^{O(m^2)}$ (modulo an $n$ factor, this is the dominant term in the complexity), noting that $\|W(x^*-z)\|_\infty \leq (m \Delta)\|x^*-z\|_1$. They then recover an optimal solution by solving the integer program $W x = b^*, x \in \{0,1\}^n$. Note that this version of the algorithm requires $W$ to be known. 

When $W$ is unknown, they show that one can replace $W$ by a proxy matrix $W'$, whose rows correspond to linearly independent gradients of $f(x) := g(Wx)$ seen so far by the algorithm. Their first observation is that the gradients of $\nabla f(x) = W\tran \nabla g(Wx)$ are linear combinations of the rows of $W$. Their second crucial observation is that for $b^* = W'x^*$, any integer solution to $W' x = b^*, x \in \{0,1\}^n$, is either optimal or has a gradient $\nabla f(x)$ outside the row span of $W'$, in which case we can add an extra row to $W'$. Thus, one can iterate the guessing procedure with $W$ replaced by $W'$ at most $m$ times before finding an optimal solution. The blowup in complexity in this setting comes from a lack of control over the coefficients appearing in $W'$. Indeed, this is precisely why they require that $g$ has an $L$-Lipschitz gradient and integral gradients on integral inputs.

We remark that this idea can be implemented more efficiently without suffering from the worse parameters of $W'$. First, we observe that the cardinality of the set
\begin{equation*}
    B_N = \{W'x : x\in \mathbb \{0, 1\}^n, \Norm{\lfloor z \rfloor - x}_1 \le N\}
\end{equation*}
can be bounded solely in $N$ and the parameters of $W$. This is because each row of $W'$ is a linear combination of rows
of $W$. Hence, $Wx = Wx'$ implies $W'x = W'x'$ and therefore $|B_N| \le O(N\Delta)^m$. Next, notice that
$B_N$ can be enumerated in time polynomial in $n$ and $|B_N|$: this follows from an induction over $n$.
To this end, for all $n'\le n + 1$ we define
\begin{equation*}
    B^{(n')}_N = \{W'x : x\in \mathbb \{0, 1\}^n, \Norm{\lfloor z \rfloor - x}_1 \le N, \text{ and } x_i = \lfloor z_i \rfloor \text{ for all } i\ge n' \} \ ,
\end{equation*}
where $B^{(n + 1)}_N = B_N$.
We now iteratively generate the sets $B^{(n')}_{N'}$ by using $B_{N'}^{1} = \{W' \lfloor z \rfloor\}$ and the recurrence
\begin{equation*}
    B_{N'}^{(n'+1)} = \begin{cases}
        B_{N'}^{(n')} \cup (B_{N'-1}^{(n')} + W'_{n'}) &\text{ if $\lfloor z_{n'} \rfloor = 0$,} \\
         B_{N'}^{(n')} \cup (B_{N'-1}^{(n')} - W'_{n'}) &\text{ if $\lfloor z_{n'} \rfloor = 1$.}
    \end{cases}
\end{equation*}
Here $W'_{n'}$ is the $n'$th column of $W'$. We note that every vector in $B_{N'-1}^{(n')}$ is generated from some $x$ with $x_{n'} = 0$ iff $\lfloor z_{n'} \rfloor = 0$. Hence, when adding (resp. removing) $W'_{n'}$ there is again a legal $x$ generating this vector (where $\Norm{\lfloor z \rfloor - x}_1$ has increased by one).

As in the algorithm of Hunkenschr\"oder et al.,
we now start with $W'$ having only the single row $\nabla f(\lfloor z \rfloor)$. Then for every element of $B_N$
we consider a corresponding integer solution $x$ (note that such an $x$ can easily be recovered in the above recurrence) and check if $\nabla f(x)$ and the rows of $W'$ are
linearly independent. If so, we add the gradient as a new row to $W'$. We repeat for at most $m$ iterations until no
new row is added. Then we return the best solution $x^*$
seen during this process.

\begin{theorem}
    Let $g: \mathbb R^m \rightarrow R$ be a convex function, let $f(x) := g(Wx)$ be accessible via a function value and gradient oracle, where $W\in\mathbb Z^{m\times n}$ is an unknown matrix with entries of absolute value at most $\Delta$. Then given an optimal solution $z$ to the continuous relaxation $\min\{f(x) : x\in [0, 1]^n\}$ with at most $m$ fractional entries, one can compute an optimal integral solution in time $n (N\Delta)^{O(m)}$.
    Here $N$ is the minimum $\Norm{z - x^*}_1$ over all optimal integer solutions $x^*$.
    In particular, when $g$ is separable convex, the running time becomes $n (m\Delta)^{O(m^2)}$.
\end{theorem}

\section{Conclusion and Open Questions}
In this paper we have demonstrated that the task of optimizing low dimensional functions over
a projection as introduced by Hunkenschr\"oder et al.~\cite{hunkenschroder2022optimizing} remains
tractable even in much more general settings than originally considered.
This creates a bridge also to other lines of work in integer optimization, such as integer programs
with few constraints~\cite{EisenbrandW2019}.

Our main result leaves open a few questions about the complexity of algorithms for
problem~\eqref{eq:problem} or the central case of $g$ being a convex function.
As mentioned before, one cannot hope to avoid a term of $\Delta^{m}$ in the running time
because of known conditional lower bounds. The necessity of the $n^m$ term or the $m^2$ exponent, however,
appears less clear.

The algorithm for integer programming by Eisenbrand and Weismantel~\cite{EisenbrandW2019},
a special case of our setting (see applications), does not require the $n^m$ term and in many cases
we can avoid it as well by merging duplicate columns of $W$. It would be nice if this term
could be removed in general, or at least in the convex case.

Related to the $m^2$ exponent, there is already a notorious question arising from~\cite{EisenbrandW2019}.
There, Eisenbrand and Weismantel gave an improved algorithm with exponent $O(m)$ instead of $O(m^2)$ for the case
that there are no upper variable bounds, but with bounds they require $O(m^2)$. It remains unclear
whether this is necessary. In our case even without upper bounds our algorithm need the exponent $O(m^2)$.
In fact, this exponent arises in several places: when guessing the support (assuming $n \approx \Delta^m$) and when guessing the projection of the tight variables $b^{(T)}$.
\bibliographystyle{plain}
\bibliography{bibliography}
\end{document}